\documentclass[prb,aps,twocolumn,showpacs,nobibnotes,epsf]{revtex4}

\usepackage{graphicx}
\usepackage{dcolumn}
\usepackage{bm}
\usepackage{SIunits}

\begin{document}

\title{Reentrant spin-glass behavior in  $TlFe_{2-x}Se_2$ with the $ThCr_2Si_2$-type structure}
\author{J. J. Ying, A. F. Wang, Z. J. Xiang, X. G. Luo, R. H. Liu, X. F. Wang, Y. J. Yan, M. Zhang, G. J. Ye, P. Cheng}
\author{X. H. Chen}
\altaffiliation{Corresponding author} \email{chenxh@ustc.edu.cn}
\affiliation{Hefei National Laboratory for Physical Science at
Microscale and Department of Physics, University of Science and
Technology of China, Hefei, Anhui 230026, People's Republic of
China\\}

\begin{abstract}
We investigated the physical properties of $TlFe_{2-x}Se_2$ single
crystals. The resistivity of $TlFe_{2-x}Se_2$ shows typical
semiconductor behavior with an activation energy of 25 meV. DC
susceptibility indicates an antiferromagnetic transition at about
450 K. Reentrant spin-glass (RSG) behavior was found at about 130 K
through DC and AC magnetic measurements. The RSG behavior suggests
the existence of a strong competition between ferromagnetic (FM) and
antiferromagnetic (AFM) interactions due to Fe deficiencies. Strong
electron-electron correlation may exist in this material and it is
possibly a candidate of parent compound for high $T_c$
superconductors.
\end{abstract}


\vskip 300 pt

\maketitle

The newly discovered iron-based superconductors have attracted much
attention in past three years\cite{Kamihara, chenxh, ren, Liu,
rotter}. All the iron-based superconductors have square planar
$Fe^{2+}$ layers. The binary chalcogenide $\alpha$-FeSe with
PbO-structure is found to be superconducting at 8 K\cite{Hsu} and
the $T_c$ could reach 37 K under high pressure\cite{Medvedev}. It is
very meaningful to investigate other materials which contain the
same Fe-Se layers. Through intercalation between the Fe-Se layers
might induce high chemical pressure and strongly increases $T_c$.
$TlFe_{2-x}Se_2$ with the $ThCr_2Si_2$-type structure consists of
Fe-Se layers which is the same with $\alpha$-FeSe except for some
iron deficiencies\cite{Klepp, Haggstrom}. The large size of Tl
cation leads to larger separation between Fe-Se layers than that in
$\alpha$-FeSe. The structure of stoichiometric $TlFe_2Se_2$ is shown
in the inset of Fig.1(a). The physical properties of this compound
have not been reported except for the antiferromagnetic transition
at 450 K revealed by the M\"{o}ssbauer measurements\cite{Haggstrom}.
It is very significant to systematically investigate its physical
properties and find out its relation to the high $T_c$
superconductors. Recent research in $Fe_{1.1}Te_{1-x}Se_x$ system
shows that spin-glass state would emerge when antiferromagnetic
state is destabilized by the Se substitution\cite{Paulose}. Spin
glasses (SGs) are the magnetic system in which the interactions
between the magnetic moments are in conflict with each other, thus
no conventional long-range order can be established\cite{Binder}.

In this paper, we investigated the physical properties of the
$TlFe_{2-x}Se_2$ single crystals. The resistivity shows typical
semiconductor-like behavior with the activated energy of 25 meV and
no superconductivity was found. Antiferromagnetic transition was
also detected by the DC susceptibility measurement at about 450 K.
Reentrant spin-glass (RSG) behavior was found at about 130 K through
DC and AC magnetic measurements. RSG transition is a well-known
phenomenon of spin glasses. The RSG system known thus far has either
long-range FM or AFM ordering above the spin glass transition
temperature\cite{Campbell}. This behavior may be related to the
competition between ferromagnetic and antiferromagnetic interactions
in the iron layers.

$TlFe_{2-x}Se_2$ single crystals were grown by directly melting the
mixture of $Tl_2Se$, Fe and Se powders in the sealed quartz tube at
the temperature of 800$^o$C. After keeping the mixture at 800$^o$C
for 2 days, the furnace was slowly cooled down to 500$^o$C at the
rate of 5$^o$C/h, then the power of furnace was shut off. Plate like
single crystals of $TlFe_{2-x}Se_2$ can be easily cleaved from the
final product. Energy-dispersive X-ray spectroscopy (EDX) shows the
iron deficiency x is about 0.35.

\begin{figure}[t]
\centering
\includegraphics[width=0.5\textwidth]{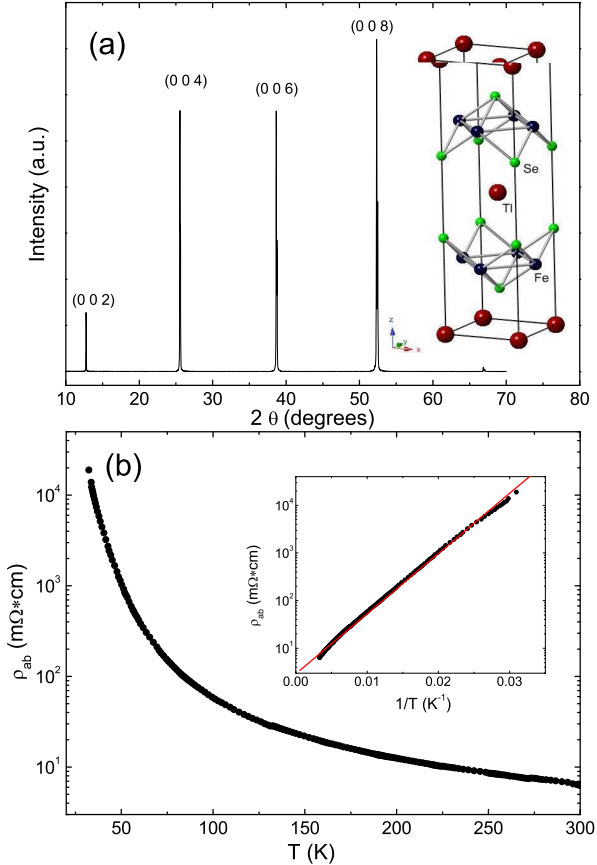}
\caption{(color online).(a): The single crystal x-ray diffraction
pattern of $TlFe_{2-x}Se_2$, Only (00$l$) diffraction peaks show up
indicating that the c axis is perpendicular to the plane of the
plate. The inset shows the crystal structure of stoichiometric
$TlFe_2Se_2$, the red, blue and green balls represent Tl, Fe and Se
ions. (b): Temperature dependence of the in-plane resistivity. The
inset shows that Log($\rho$) can be linearly fitted with 1/T. The
red line is the linear fitting result.} \label{fig1}
\end{figure}
Single crystal of $TlFe_{2-x}Se_2$ was characterized by x-ray
diffractions (XRD) using Cu $K_\alpha$ radiations. As shown in the
Fig.1(a). Only (00$l$) diffraction peaks were observed, suggesting
that the crystallographic c axis is perpendicular to the plane of
the single crystal. The lattice constant of c-axis is determined to
be 13.98 {\AA} which is slightly smaller than the reported value
14.00 {\AA}. Fig.1(b) shows the temperature dependence of the
resistivity with the electric current flowing in the ab plane of the
$TlFe_{2-x}Se_2$ single crystal. The resistivity increases rapidly
with decreasing the temperature and the room-temperature resistivity
is about 5 m$\Omega$*cm. The red line in the inset of Fig.1(b) is
the linear fit of log($\rho$) versus 1/T. The resistivity behavior
basically obeys the thermally activated behavior:
$\rho$=$\rho_0$exp($E_a$/$k_B$T), where $\rho_0$ refers to a
prefactor and $k_B$ is Boltzmann constant. The activation energy
$E_a$ was estimated to be about 25 meV from the fitting result.
However, density functional calculation shows that $TlFe_2Se_2$ has
density state at the Fermi surface, and should be metallic. This is
very different from our results\cite{Zhang}. This contradiction
could be due to the Fe deficiency in our samples or the strong
electron correlation in $TlFe_{2-x}Se_2$.

\begin{figure}[t]
\centering
\includegraphics[width=0.5\textwidth]{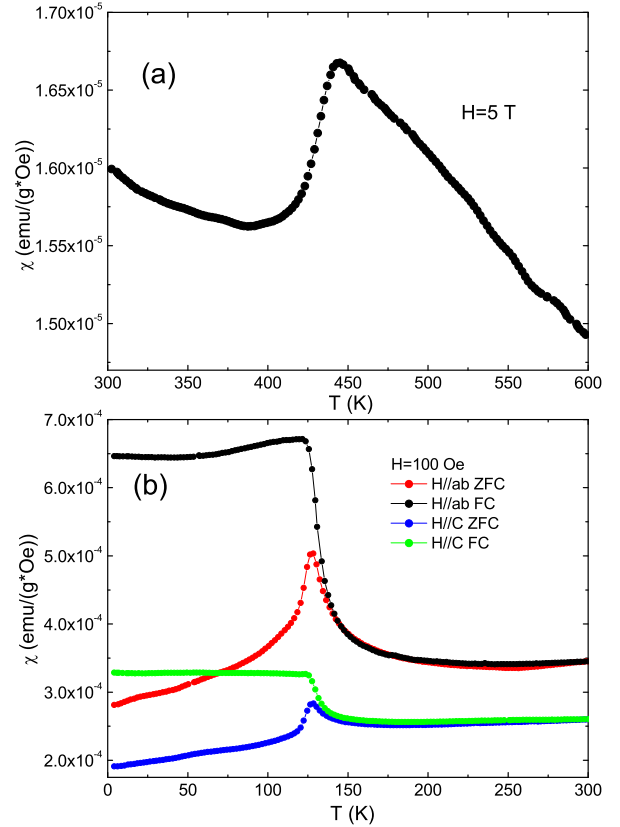}
\caption{(color online). (a): The DC magnetic susceptibility with
the applied field of 5 T along the ab plane above 300 K. (b) ZFC and
FC magnetic susceptibility with the magnetic field applied parallel
and perpendicular to the plane below 300 K, respectively. }
\label{fig2}
\end{figure}
Fig.2(a) shows the temperature dependence of the DC magnetic
susceptibility above 300 K with the magnetic field of 5 T applied
along the ab plane. The sudden drop of the magnetic susceptibility
at about 450 K indicates the antiferromagnetic ordering below 450 K.
This result is consistent with the previous report by M\"{o}ssbauer
measurements\cite{Haggstrom}. Fig.2(b) shows the anisotropic
magnetization measurements with the applied field of 100 Oe parallel
and perpendicular to the ab plane below 300 K. A pronounced
irreversibility between the zero-field-cooling (ZFC) and
field-cooling (FC) curves can be observed below 130 K. It definitely
indicates a second magnetic transition. This second magnetic
transition behaves actually like a spin-glass transition. The ZFC
curve shows a cusp-like behavior with the peak at about 130 K and
the FC curve becomes almost a constant below the reentrant
spin-glass transition temperature $T_{RSG}$ (We defined the
$T_{RSG}$ as the peak of the ac susceptibility at about 130 K shown
in Fig.3) due to the spin freezing. The magnetic susceptibility for
the field applied in the ab plane is about twice in magnitude larger
than that with the field applied perpendicular to the ab plane. It
suggests that magnetic moments in the iron plane prefer to align
along the ab plane, being consistent to the M\"{o}ssbauer
measurements and the theoretical calculation\cite{Haggstrom, Zhang}.
The susceptibility almost does not change with decreasing the
temperature above 200 K due to the nature of the long-ordering AFM
state. With the temperature approaching $T_{RSG}$, the
susceptibility increases very rapidly with the magnetic field
applied both along and perpendicular to the ab plane. It might be
due to an increase of the ferromagnetic interactions in the iron
layers around $T_{RSG}$. Actually, the ferromagnetic interactions
start to emerge at the temperature much higher than $T_{RSG}$ as the
susceptibility starts to increase at the temperatures much higher
than $T_{RSG}$.

Fig.3(a) and (b) show the real part of ac magnetic susceptibility
with the frequency of 1 Hz with the ac magnetic field ($H_{ac}$ =
3.8 Oe) parallel and perpendicular to the ab plane, respectively.
The real part of ac magnetic susceptibility slightly increases with
decreasing the temperature at high temperature and shows a peak at
about 130 K. The sharp drop in ac susceptibility  below $T_{RSG}$
indicates the spin freezing below $T_{RSG}$. $T_{RSG}$ shifts to
higher temperature with increasing the frequency to 1500 Hz as shown
in the inset of Fig.3(a) and (b). This behavior strongly supports
the glassy nature of the magnetic state in this system. Larger value
of $\chi_{ab}'$ than $\chi_c'$ also indicates that the magnetic
moments prefer to align along the ab plane.
\begin{figure}[t]
\centering
\includegraphics[width=0.5\textwidth]{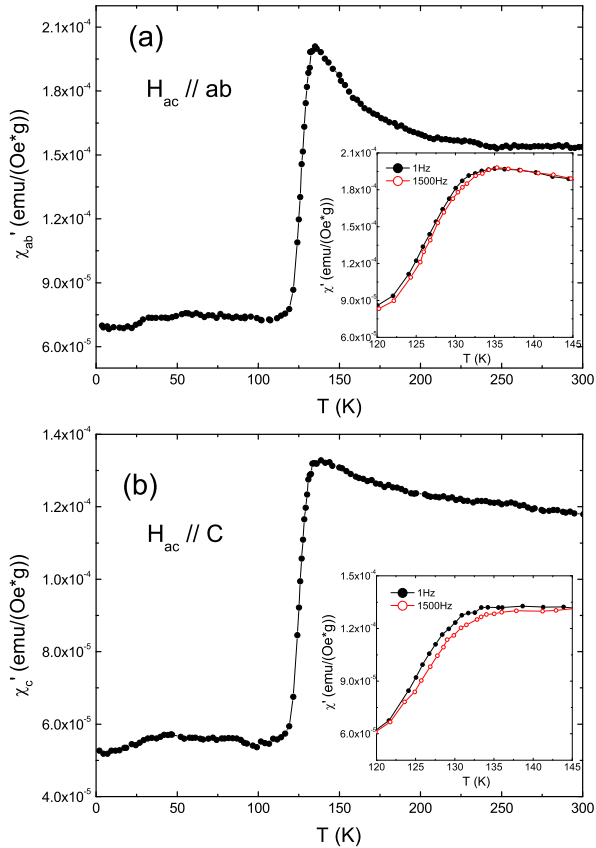}
\caption{(color online). The real part of ac susceptibility with the
frequency of 1 Hz with the ac field applied along and perpendicular
to the plane (a) and (b), respectively. The insets show the
frequency shift around the RSG transition temperature. The red line
is the ac susceptibility with the frequency of 1500 Hz.}
\label{fig3}
\end{figure}

\begin{figure}[t]
\centering
\includegraphics[width=0.5\textwidth]{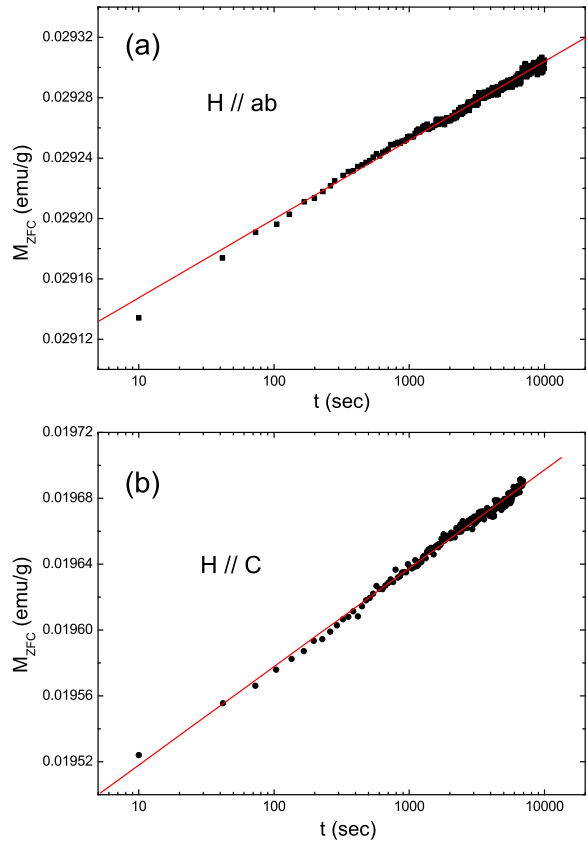}
\caption{(color online). The long-time relaxation behavior of
$M_{ZFC}$(t) for the magnetic field applied parallel (a) and
perpendicular (b) to the ab plane. The red line is the linear
fitting of the $M_{ZFC}$ with log (t).} \label{fig4}
\end{figure}

\begin{figure}[t]
\centering
\includegraphics[width=0.45\textwidth]{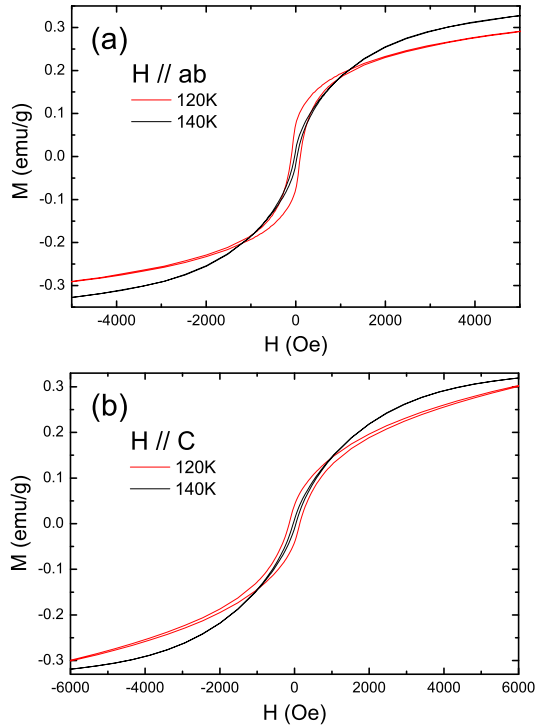}
\caption{(color online). Isothermal magnetization hysteresis loop
with the magnetic field applied parallel (a) and perpendicular (b)
to the ab plane. The red hysteresis curve is at 120 K below
$T_{RSG}$ and the black one is at 140 K above $T_{RSG}$.}b
\label{fig5}
\end{figure}

\begin{figure}[t]
\centering
\includegraphics[width=0.45\textwidth]{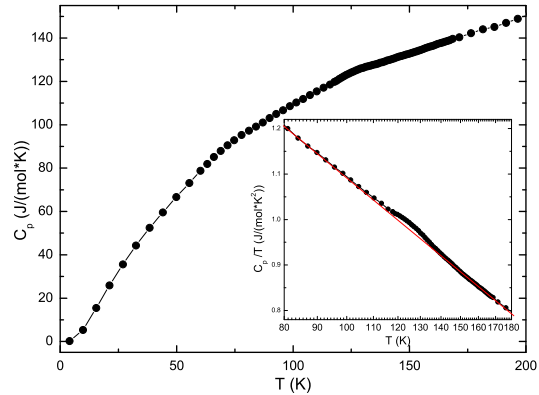}
\caption{(color online). Temperature dependence of the heat
capacity. The inset is the enlarged area around $T_{RSG}$ of
$C_p/T$. The red line is the contrast line for us to easily detect
the broad hump around $T_{RSG}$.} \label{fig6}
\end{figure}

To further investigate the RSG behavior, we measured the long-time
relaxation of the DC magnetization. Fig.4 (a) and (b) show the
long-time relaxation behavior of $M_{ZFC}$(t) for the magnetic field
applied parallel and perpendicular to the ab plane, respectively.
The experiment process is as follows. The crystal was cooled without
magnetic fiels (the ZFC process) from the room temperature to 4
K($<$$T_{RSG}$), then the magnetic field of 100 Oe was applied
parallel or perpendicular to the plane. As soon as the field was
applied, the $M_{ZFC}$ was measured as a function of time t. As
shown in Fig.4(a) and (b),  it is found that the $M_{ZFC}$ slowly
increases almost linearly with the log (t). The long-time relaxation
evidences an important role of the frustration as observed in the SG
region of $La_{1-x}Sr_xCoO_3$ perovskite compounds\cite{masayuki,
Nam}. This long-time relaxation behavior in the low temperature
region is consistent with the RSG state\cite{Joonghoe, Jonason,
Mathieu}.

Fig.5(a) and (b) show the magnetization hysteresis curves for the
magnetic field applied parallel and perpendicular to the ab plane,
respectively. An S-shape M-H Loop is observed at 120 K which is a
little lower than $T_{RSG}$. This behavior is typical in SG system.
The magnitude of coercive fields are 100 Oe and 150 Oe for the
magnetic field applied parallel and perpendicular to the plane,
respectively. While above $T_{RSG}$, only unobvious loop can be
found. This behavior may be attributed to the competition between
ferromagnetic and antiferromagnetic interactions. Above $T_{RSG}$
antiferromagnetic is much stronger than ferromagnetic interactions.
While below $T_{RSG}$, ferromagnetic interaction is comparable to
the antiferromagnetic interaction, some part of the system may form
domains. Thus due to the blocking of the domain wall motion similar
to ferromagnetic materials M-H loop occurs. The competition of
ferromagnetic interaction and the antiferromagnetic interaction may
be the main cause of the RSG transition. The s-shape M-H loop above
$T_{RSG}$ also indicate that the short-range ferromagnetic
interaction already exist above $T_{RSG}$ and the strong fluctuation
in this system.

In order to further investigate the spin-glass transition at around
130 K, we performed heat capacity measurement at the temperature
range from 4 K to 200 K. Fig.6 shows the result of the heat capacity
for $TlFe_{2-x}Se_2$. A very weak broad hump of $C_p/T$ extending
over a wide temperature range can be found around $T_{RSG}$ as shown
in the inset of Fig.6. It confirms that the transition around 130 K
is the bulk property for the $TlFe_{2-x}Se_2$. Such a broad hump of
the heat capacity is often observed in spin-glass
system\cite{brodale, Martin}.

The spin-glass behavior has been reported in $FeSe_{1-x}Te_x$
system\cite{Paulose}. A Similar SG state is found in the
$TlFe_{2-x}Se_2$ with $ThCr_2Si_2$ structure. The RSG state in the
$TlFe_{2-x}Se_2$ is mainly attributed to arise from the competing
interactions between the FM and AFM phases. M\"{o}ssbauer
measurement shows that the antiferromagnetic transition temperature
is greatly suppressed and the transition width broaden with
increasing the iron vacancy\cite{Haggstrom2}. It suggests that the
iron vacancies would strongly destroy the antiferromagnetic ground
state of the the Fe ion layers, and the fluctuation is induced
between the neighbor Fe ions around the Fe vacancies. The spin-glass
like behavior was also found in the isostructure compounds
$ACuFeS_2$ (A=K, Rb, Cs), they can be represented as micromagnets at
low temperature and the superexchange interactions between spins
within a cluster depend on the distribution of iron
ions\cite{Oledzka}. In our system, the Fe vacancies may separate the
Fe spins and Fe cluster can be formed, consequently lead to the spin
glass state. All these results indicate that the Fe ion deficiencies
may be the origin of the RSG behavior in $TlFe_{2-x}Se_2$. The
semiconductor-like resistivity is quite different to the metal state
for the material without iron deficiencies calculated by density
functional calculation. The typical semiconductor resistivity
behavior is also found in $ACuFeS_2$. The non-linear M-H curve
indicates the strong fluctuation in the iron layers. The high
antiferromagnetic transition temperature and insulator ground state
is very similar to the parent compounds of cuprates. Strong electron
correlation may exist in this kind of material and it is possibly a
candidate of parent compounds for high Tc superconductors. In order
to explore superconductivity in this material, the problem is how to
overcome the Fe deficiency in the iron-layers.

In conclusion, we systematically investigated the physical
properties of the $TlFe_{2-x}Se_2$ single crystals. The resistivity
shows typical semiconductor behavior, and no superconductivity was
found at the temperature cooling down to 2 K. At 450 K,
antiferromagnetic transition was observed by DC magnetic
measurement, being consistent with the early M\"{o}ssbauer result.
Both DC and AC magnetic measurements show the RSG transition at the
temperature around 130 K. The RSG state could arise from the
competition between the FM and AFM interactions. Such competing
interactions may be due to the iron deficiencies, and such
deficiencies induce the fluctuation between the neighbor Fe ions
around the Fe vacancies. The heat capacity shows a very small and
broad hump with no resemblance to a $\lambda$-type peak around the
$T_{RSG}$ . This material might be a candidate of parent compounds
for high Tc superconductors with the same structure to the 122 phase
in iron based superconductors.

This work is supported by the Nature Science Foundation of China,
and by Ministry of Science and Technology and by Chinese Academy of
Sciences.

\end{document}